\documentstyle[pre,aps,multicol,epsfig]{revtex}

\begin{document}
\draft
\title{Experimental vs. Numerical Eigenvalues of a Bunimovich 
       Stadium Billiard --\\  A Comparison
      }
\author{H. Alt$^{1}$\footnote{Present address: Kl\"ockner Pentaplast, D-56412 
        Heiligenroth, Germany }, C. Dembowski$^{1}$, 
        H.-D. Gr\"af$^{1}$, R. Hofferbert$^{1}$,
        H. Rehfeld$^{1}$, A. Richter$^{1,2}$ and C. Schmit$^{3}$
       }

\address{$^{1}$ Institut f\"ur Kernphysik, Technische Universit\"at Darmstadt,
         D-64289 Darmstadt, Germany \\                                  
         $^{2}$ Wissenschaftskolleg zu Berlin, D-14193 Berlin, Germany\\
         $^{3}$ Institute de Physique Nucl\'eaire,
         F-91406 Orsay, France\\        
        }

\date{\today}
\maketitle
\begin{abstract}
We compare the statistical properties of eigenvalue sequences for a 
$\gamma=1$ Bunimovich stadium billiard. The eigenvalues have been obtained by 
two ways: one set results from a measurement of the eigenfrequencies of a
superconducting microwave resonator ({\it real} system) and the other set is 
calculated numerically ({\it ideal} system). The influence of the mechanical  
imperfections
of the {\it real} system in the analysis of the spectral fluctuations and in 
the length spectra compared to the exact data of the {\it ideal} system are 
shown. We also discuss the influence of a family of marginally stable orbits, 
the bouncing ball orbits, in two microwave stadium billiards with 
different geometrical dimensions.
\end{abstract}
\begin{multicols}{2}
\narrowtext

\section{Introduction}
\label{intruduction}
Quantum manifestations of classical chaos have received much attention in 
recent years \cite{gutzwiller} and for the semiclassical quantization of 
conservative chaotic systems, two-dimensional billiard systems provide a very 
effective tool \cite{mcdonald,berry_87}. Due to the conserved energy of the
ideal particle propagating inside the billiard's boundaries with specular
reflections on the walls, the billiards belong to the class of Hamiltonian
systems with the lowest degree of freedom in which chaos can occur and this
only depends on the given boundary shape. Such systems are in particular 
adequate to study the behavior of the particle in the corresponding quantum 
regime, where spectral properties are completely described by the stationary 
Schr\"odinger equation. The spectral fluctuations properties of such systems 
were investigated both analytically and numerically. It has been found that 
these properties coincide with those of the ensembles of random matrix theory 
(RMT) having the proper symmetry \cite{mehta,bohigas_91} if the given system 
is classically non-integrable. For time-reversal invariant systems, to which 
group the here investigated billiards belong, the relevant ensemble is the 
Gaussian orthogonal ensemble (GOE).

In the last decades this subject was dominated by theory and numerical 
simulations. About seven years ago experimentalists have found effective 
techniques to simulate quantum billiard problems with the help of macroscopic 
devices. Due to the equivalence of the stationary Schr\"odinger equation and 
the classical Helmholtz equation in two dimensions one is able to model the 
billiard by a similarly shaped electromagnetic resonator 
\cite{sridhar,stoeckmann,graefprl_92}.

\begin{figure} [hbt]
\centerline{\epsfxsize=8.6cm
\epsfbox{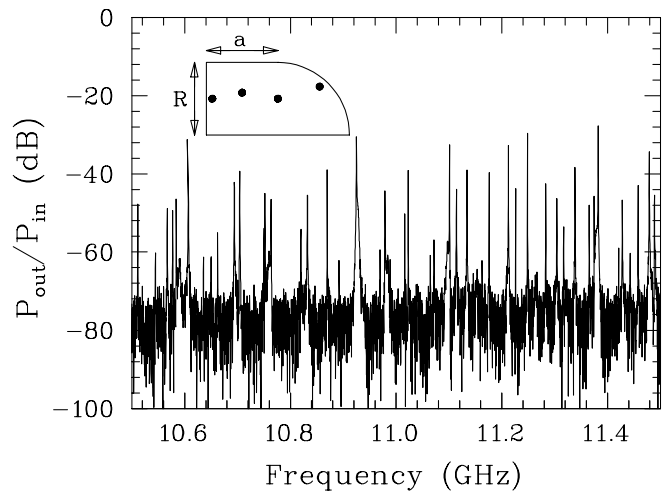}}
\vspace*{2.5ex}
\caption{Typical transmission spectrum of the superconducting $\gamma=1$ 
stadium billiard fabricated from niobium in the range between 10.5 and 
11.5 GHz taken at 4.2 K. The signal is given as the ratio of output power to 
input power on a logarithmic scale. The inset illustrates the shape of the 
resonator and the positions of the antennas.}
\end{figure}

Theoretical predictions assume to have always an {\it ideal} system with a
perfect geometry whereas experiments have been performed with  {\it real} 
systems. Real microwave cavities -- in particular the here used superconducting 
ones -- are usually not machined by the most accurate technique, e.g. milling 
from a solid block. They are cut from niobium sheet material which is shaped 
and afterwards electron beam welded. Finally they are chemically etched to 
clean their inner surface. Their final inner geometry is not accessible for a 
direct measurement so that their exact geometric properties are not known. In 
the case of the investigated Bunimovich stadium billiard \cite{bunimovich}, 
where $\gamma=a/R=1$ (see inset of Fig.~1), the following properties are 
especially crucial: The radius of curvature of the boundary does not change 
abruptly but smoothly at the transition from the straight $a$ to the circular 
section of the boundary \cite{alonso}. Another point is that the  angle at the 
corners is not exact 90 degrees. 

The paper is about the question to which extent a comparison of experimental 
data with theoretical predictions for such billiards is meaningful. Therefore 
we want to compare a numerical simulation (calculation of eigenvalues) for a 
$\gamma=1$ stadium billiard with a measurement of a real superconducting 
microwave cavity (measurement of eigenvalues) by studying the statistical 
properties of the two sequences of eigenvalues.

The paper is organized as follows. In Sec.~\ref{experiment} the experimental
set-up and the measurement of the eigenfrequencies are described and in 
Sec.~\ref{numeric} the numerical calculations. The comparison of both sets of 
eigenvalues is shown in Sec.~\ref{results} by analyzing their spectral 
fluctuations.

\section{Experiment}
\label{experiment}

Experimentally we have investigated a two-dimensional microwave cavity which 
simulates a two-dimensional quantum billiard. Here we present results based on
measurements using a superconducting niobium cavity, having the shape of a 
quarter Bunimovich stadium billiard. The billiard has been desymmetrized to 
avoid superpositions of several independent symmetry classes 
\cite{bogomolny_88}. Its inner dimensions are $a=R=20$ cm corresponding to 
$\gamma=a/R=1$, see inset of Fig.~1, and it has a height of $d=0.7$ cm, that 
guarantees a two-dimensionality up to a frequency of 
\mbox{$f_{max}=c/2d=21.4$ GHz} ($c$ denotes the speed of light). 

As in previous investigations \cite{alt_3d,alt_coupling} the measurement of 
the $\gamma=1$ stadium billiard has been carried out in a LHe-bath cryostat. 
This experimental set-up is very stable concerning temperature and pressure 
fluctuations. The cavity has been put into a copper box which was covered by 
liquid helium, so that a constant temperature of 4.2 K inside the resonator 
was guaranteed during the whole measurement. The box has also been evacuated 
to a pressure of $10^{-2}$ mbar to eliminate effects of the dielectric gas 
inside the cavity. We were able to excite the cavity in the frequency range of 
\mbox{$45$ MHz $< f< 20$ GHz} in 10 kHz-steps using four capacitively coupling 
dipole antennas sitting in small holes on the niobium surface, see inset of 
Fig.~1. These antennas penetrated only up to a maximum of 0.5 mm into the 
cavity to avoid perturbations of the electromagnetic field inside the 
resonator. Using one antenna for the excitation and either another one or the 
same one for the detection of the microwave signal, we are able to measure the 
transmission as well as the reflection spectra of the resonator using an 
HP8510B vector network analyzer. In Fig.~1 a typical transmission spectrum of 
the billiard in the range between 10.5 and 11.5 GHz is shown. The signal is 
given as the ratio of output power to input power on a logarithmic scale. The 
measured resonances have quality factors of up to 
\mbox{$Q=f/\Delta f \approx 10^7$} and signal-to-noise ratios of up to 
$S/N\approx 60$ dB which made it easy to separate the resonances from each 
other and detect also weak ones above the background. As a consequence of using 
superconducting resonators, all the important characteristics like 
eigenfrequencies and widths can be extracted with a very high accuracy 
\cite{altprl_95,altnucphys_93,altplb_96}. A detailed analysis of the raw 
spectra yielded a total number of 955 resonances up to 20 GHz. To reduce the 
possibility of missing certain modes with a rather weak electric field vector 
at the position of the antennas, the measurements were always performed with 
different combinations of antennas. Thereby the number of missed modes is 
dramatically reduced below three to five in a typical case of a measurement of 
about one thousand eigenfrequencies. For the $\gamma=1$ stadium billiard 
investigated here the 955 detected eigenmodes agrees exactly with the expected 
number calculated from particular geometry of the cavity with the help of 
Weyl's formula (see Sec.~\ref{results} below). The measured sequence of 
frequencies extracted from the experimental spectra forms the basic set of the 
statistical investigations in Sec.~\ref{results}.

\section{Numerical Calculations}
\label{numeric}
\subsection{ Theory}
The problem that we have to solve is to find the eigenvalues of the Dirichlet 
problem in a billiard, i.e. to solve the following equations:
\begin{equation}   
\Delta \Psi (\vec r) +k^2 \Psi (\vec r) = 0 \mbox{ for } \vec r \in {\cal D}  
\label{eq1}
\end{equation}
\begin{equation}          
\Psi (\vec r)=0  \mbox{ for } \vec r \in \partial {\cal D}  
\label{eq2}
\end{equation}

In order to solve these equations we search for a solution in form of an 
expansion of regular Bessel functions, so that we write

\begin{equation}  
\Psi (\vec r)= \Psi (r,\theta)
             = \sum_{l=-L}^{L} a_l J_l(k r) e^{il\theta} 
\label{eq3}
\end{equation}

This expansion is obviously a solution of the first equation, so that we have 
to solve it under the boundary conditions Eq.~(\ref{eq2}). On the billiard 
boundary, which may be parameterized by the curvilinear abscissa $s$, the 
expansion Eq.~(\ref{eq3}) becomes

\begin{equation} 
\Phi (s)= \Psi (r(s),\theta(s))
        = \sum_{l=-L}^{L} a_l J_l(k r(s)) e^{il\theta(s)} 
\label{eq4}
\end{equation}
which is a periodic function of $s$. This periodic function may be decomposed
in a Fourier series whose coefficients are given by

\begin{equation}  
C_n(k)={1 \over{2\pi}} \int_0^{\cal L} ds \, \Phi (s) e^{-2 i n \pi
 s/{\cal L}}
      = \sum_{l=-L}^{L} a_l C_{n,l}(k)    
\label{eq5}
\end{equation}
with
\begin{equation}
C_{n,l}(k)=\int_0^{\cal L} ds \,  e^{-2 i n \pi s/{\cal L}}
           J_l(k r(s)) e^{il\theta(s)}
\label{eq6}
\end{equation}

To satisfy the boundary condition Eq.~(\ref{eq2}), one may impose the 
equivalent conditions that all Fourier coefficients $C_n(k)$ are zero, at least
for $-L \leq n \leq L$, so that we are left with the following linear system 
in the $a_l$:

\begin{equation}        
\sum_{l=-L}^{L} a_l C_{n,l}(k)=0  , -L \leq n \leq L     
\label{eq7}
\end{equation}

For this homogeneous system to have a non trivial solution, one has the
following set of equations:

\begin{equation}      
D(k)=det[C_{n,l}(k)]=0  
\label{eq8}
\end{equation}
Thus one is left with the problem of constructing the matrix $C_{n,l}(k)$ and
finding the zeroes of its determinant $D(k)$. Before going further, we would
like to point out the advantages of this method as compared to the well known
collocation method. In our method, one may vary independently the number of
boundary points used to evaluate the integrals $C_{n,l}(k)$ and the number of 
partial waves used in the expansion, whereas this is not so in usual boundary 
methods. (For other methods we refer the reader e.g. to Refs. 
\cite{mcdonald,sieber1}). Furthermore, it appears clearly that if the billiard 
is close to a circle, the matrix is nearly diagonal so that its determinant is
easy to compute numerically, whereas this is not true for plane wave 
decompositions. Finally, in contrast with the Green's method, one may always 
deal with real matrices by decomposing on sine and cosine rather than on 
exponentials (for time reversal invariant systems). However, one should note
that this method does not work for billiards whose boundary consists of
several distinct curves (for instance Sina\"{\i} billiards).

\subsection{Practical}
In practice, the application of the above algorithm depends on the problem one
has to solve. For instance, a particular choice of the origin of the 
coordinates may simplify notably the evaluation of the matrix: For the stadium,
a good choice would be the symmetry center of the billiard, so that one may 
separate easily different symmetry classes. With this choice, the expansion 
given in Eq.~(\ref{eq3}) become for odd-odd symmetry:
\begin{equation}  
\Psi (\vec r)= \Psi (r,\theta)
             = \sum_{n=1}^{N} a_n J_{2n}(k r) sin({2n\theta}) 
\label{eq9}
\end{equation}
Therewith one tabulates the function $D(k)$ in an interval of $k$ such that 
the number of partial waves needed is constant in this interval:

\begin{equation}  
k_{min} \leq k \leq k_{max}  
\label{eq10}
\end{equation}
with
\begin{equation}  
k_{min} R_{max}=L \mbox{ and } k_{max} R_{max}=L+2 
\label{eq11}
\end{equation}
In this equation, $R_{max}$ is the greatest distance of the boundary from the 
origin, in our case $R_{max}=R+a$. The number of coefficients $a_n$ may then be 
taken as 
\begin{equation}
N=L/2 +dN
\end{equation} 
where $dN$ typically ranges from 0 to 3 or 4. Usually, the positions of the 
eigenvalues depend very little on this parameter.

The integrals
\begin{equation}
C_{n,m}(k)=\int_0^{\cal L} ds \,  sin({n \pi s/{\cal L}})
           J_{2m}(k r(s)) sin({2m\theta(s)})
\label{eq12}
\end{equation}
are evaluated using points regularly spaced along the boun\-dary. As soon as 
their spacing is such that the fastest varying phase in Eq.~(\ref{eq12}) 
changes by less than $\pi$ in a given step, the evaluation of these integrals 
is accurate enough to give the position of the zeroes of $D(k)$. In other 
words, the step $\Delta s$ used in evaluating Eq.~(\ref{eq12}) should be such 
that the following condition is verified:
\begin{equation}
 {{N \pi \Delta s}\over{\cal L}}+{{2 N \Delta s}\over {d_{min}}} \le \pi \ ,
\label{eq13}
\end{equation}
where $d_{min}$ is the smallest distance of the boundary from the origin, in 
our case $R$. With these ingredients, the precise computation of the levels of 
a rather regular billiards is fast, and does not require powerful computers. 
Typically, for the stadium, the computation of the 1000 first levels takes a 
few hours run on a personal computer, the dimension of the matrices used being 
smaller than 100. The precision obtained is much better than $1/100{th}$ of 
the average spacing, which is enough for our purpose.

\section{Spectral fluctuations}
\label{results}
In the following section we want to discuss the statistical properties of the 
data obtained for the $\gamma=1$ stadium billiard as described in
Secs.~\ref{experiment} and \ref{numeric}. After a short summary of the general 
concepts of analyzing spectral fluctuations in Sec.~\ref{results_a}, we present
the results for the comparison of the  measured and calculated data in 
Sec.~\ref{results_b}. Finally, in Sec.~\ref{results_c} we compare the 
influence of a special family of orbits in different Bunimovich stadium 
billiards.

\subsection{Theoretical background}
\label{results_a}
From the measured resp. numerically simulated eigenvalue sequences (``stick 
spectrum'') the spectral level density \mbox{$\rho (k)=\sum_i \delta (k-k_i)$} 
is calculated ($k$ is the wave number, \mbox{$k=2\pi/c\cdot f$}) and a 
staircase function \mbox{$N(k)=\int \rho (k')dk'$} is constructed which 
fluctuates around a smoothly varying part, defined as the average of $N(k)$. 
Usually this smooth part $N^{smooth}(k)$ is related to the volume of the 
classical energy-allowed phase-space. For the 2D billiard at hand with 
Dirichlet boundary conditions it is given by the Weyl-formula 
\cite{weyl1,weyl2}
\begin{equation}
N^{Weyl}(k)\approx\frac{A}{4\pi}k^2-\frac{C}{4\pi}k+{\rm const.}
\label{weylformula}
\end{equation}
where $A$ is the area of the billiard and $C$ its perimeter. The constant term 
takes curvature and corner contributions into account. Higher order terms 
\cite{baltes} are not relevant for the present analysis. The remaining 
fluctuating part of the staircase function 
\mbox{$N^{fluc}(k)=N(k)-N^{Weyl}(k)$} oscillates around zero. While 
Eq.~(\ref{weylformula}) does not contain any information regarding the 
character of the underlying classical dynamics of the system, the fluctuating 
part does.

In order to perform a statistical analysis of the given eigenvalue sequence 
independently from the special size of the billiard, the measured resp. 
calculated spectrum is first unfolded \cite{bohigas_84}, i.e. the average 
spacing between adjacent eigenmodes is normalized to one, using 
Eq.~(\ref{weylformula}). This proper normalization of the spacings of the 
eigenmodes then leads to the nearest neighbour spacing distribution $P(s)$,
from now on called NND, the probability of a certain spacing $s$ between two 
adjacent unfolded eigenfrequencies. To avoid effects arising from the bining 
of the distribution, we employ here the cumulative spacing distribution 
$I(s)=\int P(s) ds$.  

To uncover correlations between nonadjacent resonan\-ces, one has to use a
statistical test which is sensitive on larger scales. As an example we use the 
number variance $\Sigma^2$ originally introduced by Dyson and Mehta for studies
of equivalent fluctuations of nuclear spectra \cite{mehta,dyson}. The 
$\Sigma^2$-statistics describes the average variance of a number of levels 
$n(L)$ in a given interval of length $L$, measured in terms of the mean level 
spacing, around the mean for this interval, which is due to the unfolding equal
to $L$,
\begin{equation}
\Sigma^2(L)=\bigg\langle\left(n(L)-\langle n(L)\rangle_L\right)^2\bigg\rangle_L=
\langle n^2(L)\rangle_L-L^2  .
\label{sigma2formula}
\end{equation}

Furthermore, to characterize the degree of chaoticity in the system, the 
spectra are analyzed in terms of a statistical description introduced in a 
model of Berry and Robnik \cite{berryrobnik} which interpolates between the 
two limiting cases of pure Poissonain and pure GOE behavior for a classical 
regular or chaotic system, respectively. The model introduces a mixing 
parameter $q$ which is directly related to the relative chaotic fraction of 
the invariant Liouville measure of the underlying classical phase space in 
which the motion takes place ($q=0$ stands for a regular and $q=1$ for a 
chaotic system).

The chaotic features of a classical billiard system are characterized by the 
behavior of the orbits of the propagating point-like particle. The quantum 
mechanical analogue does not know orbits anymore but only eigenstates, i.e. 
wave functions and corresponding eigenenergies. Thus, the individual and 
collective features of the eigenstates must reflect the behavior of the 
classical orbits. The semiclassical theory of Gutzwiller \cite{gutzwiller} 
assumes that a chaotic system is fully determined through the complete set of 
its periodic orbits. The influence of the isolated periodic orbits of a 
billiard is most instructively displayed in the Fourier transformed (FT) 
spectrum of the eigenvalue density $\rho^{fluc}(k)=dN^{fluc}(k)/dk$, i.e.
\begin{equation}
\tilde{\rho}^{fluc}(x)=\int_{k_{min}}^{k_{max}} e^{ikx}\left[\rho (k)-
\rho^{Weyl}(k)\right]dk , 
\label{ft}
\end{equation}
with $[k_{min},k_{max}]$ being the wave number interval in which the data are 
taken.

\subsection{Results}
\label{results_b}
In this section we present the results of the analysis of our data using the
techniques described in Sec.~\ref{results_a}. We start with a direct 
one-to-one comparison of both, experimentally and numerically obtained data 
sets, and perform first a comparison with the sequences of the unfolded 
eigenvalues. To obtain the unfolded eigenvalues, we fit the Weyl-formula, 
Eq.~(\ref{weylformula}), onto our measured resp. numerically simulated spectral 
staircase $N(k)$. Doing this, one obtains the following parameter of 
Eq.~(\ref{weylformula}) for the measured data: an area  
$A_{fit,exp}=(710.52\pm4.12){\rm\ cm}^2$ and a perimeter 
$C_{fit,exp}=(113.68\pm2.16){\rm\ cm}$, which are very close to the 
design values $A_{design}=714.15{\rm\ cm}^2$ and 
$C_{design}=111.41{\rm\ cm}$. Within the given uncertainties the respective 
coefficients $A$ and $C$ agree fairly well. From the numerical data which were 
obtained from a $\gamma=1$ stadium billiard with area $A=4\pi$ the coefficient 
$C$ can be calculated. The fit of the Weyl-formula to the numerical data 
yields values for coefficients $A$ and $C$ within a few per mille of the 
calculated coefficients. From this we conclude that due to the not well known 
mechanical imperfections of the billiard the uncertainties in the average 
properties of the spectrum expressed through the coefficients $A$ and $C$ of 
the Weyl-formula are larger in the experiment than in the numerical simulations.

With the unfolding indicated above the different geometrical properties of the
{\it real} and {\it ideal} billiards are removed and a direct comparison of 
the first 770 eigenvalues becomes possible. Computing the difference of the 
experimental and the numerical data, $\epsilon_{exp}-\epsilon_{num}$, and 
plotting this difference over the unfolded experimental eigenvalues, the upper 
part of Fig.~2 is obtained. The curve fluctuates around zero, which is an 
indication for complete spectra, i.e. no mode is missing. The appearing 
oscillations in the curve are random and reflect the fact of how accurate a 
certain eigenfrequency could be determined experimentally. If only one 
eigenvalue in the measured  sequence is artificially removed, the curve shows 
a clearly observable step of height 1 at the position where the eigenvalue was 
dropped. This is demonstrated in the lower part of Fig.~2, where the 
$427^{th}$ mode has been removed in the experimental spectrum. More information 
can not be extracted from this direct comparison since systematic errors and 
imperfection have been removed through the unfolding procedure. Therefore we 
want to concentrate in the following on the statistical properties of the 
presented complete sequences of data.

\begin{figure} [hbt]
\centerline{\epsfxsize=8.6cm
\epsfbox{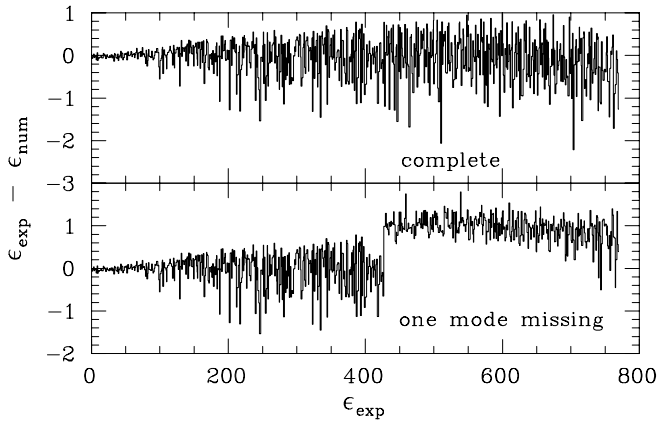}}
\vspace*{2.5ex}
\caption{In the upper part the difference of the unfolded eigenvalues
$\epsilon_{exp}-\epsilon_{num}$ is plotted over the unfolded eigenfrequencies
$\epsilon_{exp}$ of the experimental data. In the lower part the same curve is 
shown, but with the $427^{th}$ mode artificially removed from the experimental
spectrum.}
\end{figure}

In the next step, we extract the fluctuating part of the staircase function 
$N^{fluc}(k)=N(k)-N^{Weyl}(k)$ from the measured resp. calculated eigenvalue 
sequences (Fig.~3). As expected oscillations around zero are seen, also 
indicating no missing modes. 

In the upper part of Fig.~3 a strong enhancement of the amplitude of the 
fluctuations as a function of wave number $k$ -- a characteristic feature of 
regular systems -- is observed as well as a periodic gross structure with 
spacing of \mbox{$15.7\ {\rm\ m}^{-1}$} (750 MHz). This behavior of 
$N^{fluc}(k)$ is caused by the well known family of marginally stable periodic 
orbits, which bounce between the two straight segments of the billiard, the 
so-called bouncing ball orbits (bbo), with length $2r$. The cumulative level 
density $N(k)$ shows therefore periodic oscillation with a fixed period of 
$2\pi /2R=15.7$ m$^{-1}$ around the value given by the Weyl-formula. These 
observations can be described \cite{sieber} by the semiclassical expression of 
the contribution of the bbo to the spectrum which reads
\begin{equation}
N^{bbo}(k)=\frac{a}{r}\left(\sum_{1\le n\le X}\sqrt{X^2-n^2}-\frac{\pi}{4}X^2
+\frac{1}{2}X\right)
\label{n_bbo}
\end{equation}
with $X=(kr)/\pi$. This formula (for a generalization of it in three 
dimensions see \cite{alt_3d}) reproduces the mean behavior of the experimental 
data as shown in the upper part of Fig.~3. After subtraction of this smooth 
correction in addition to the expression of Weyl the proper fluctuating part 
of the level density is obtained, which is plotted in the lower part of 
Fig.~3. Naturally the same result is obtained for the data from the numerical 
calculations.

\begin{figure} [hbt]
\centerline{\epsfxsize=8.6cm
\epsfbox{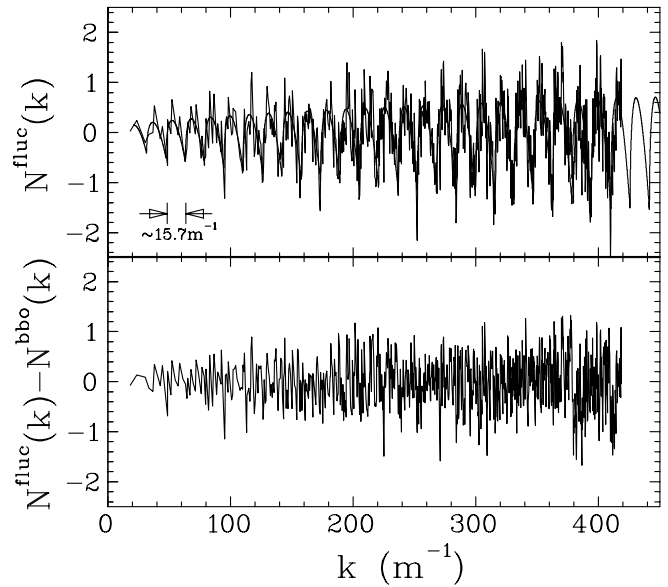}}
\vspace*{2.5ex}
\caption{The histogram in the upper part displays the fluctuating part 
$N(k)-N^{Weyl}(k)$ of the staircase function. The full line shows the 
semiclassical prediction for the bouncing ball orbits (bbo) according to 
Eq.~(\ref{n_bbo}). The lower part shows $N^{fluc}(k)-N^{bbo}(k)$, the 
fluctuating part after subtracting the contribution of the bbo.}
\end{figure}
 
To determine the degree of chaoticity of the investigated $\gamma=1$ stadium
billiard we next calculate the cumulative nearest neighbour spacing 
distribution $I(s)$ for the fluctuating part of the staircase function 
corrected by the Weyl and the bbo terms. In Fig.~4 the dashed curve shows the 
density for a certain spacing $s$ of two adjacent unfolded eigenmodes for the 
experimental and the numerical data. Beside the data also the two limiting 
cases, the Poisson- and the GOE-distribution, are displayed. Using the ansatz 
of Berry-Robnik one obtains a mixing parameter $q=0.97\pm 0.01$ for the 
experimental and $q=0.98\pm 0.02$ for the numerical data.  As stated in 
\cite{bunimovich} the Bunimovich stadium billiard should be fully chaotic, 
which is expressed through $q$ being very close to unity within the 
uncertainties of the fitting procedure.

\begin{figure} [hbt]
\centerline{\epsfxsize=8.6cm
\epsfbox{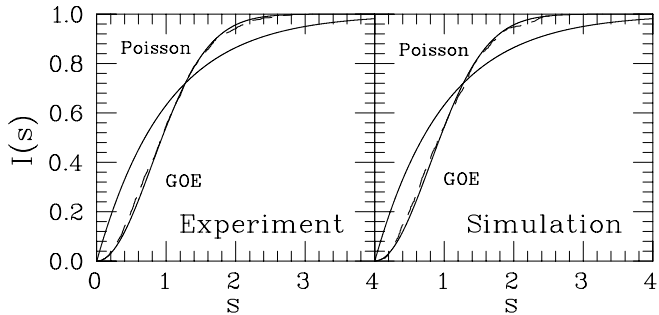}}
\vspace*{2.5ex}
\caption{Cumulative nearest neighbour spacing distribution for the $\gamma=1$ 
stadium billiards. The dashed curve corresponds to the experimental (left side)
respectively the numerical (right side) data. Also the two limiting cases 
(Poisson and GOE) are displayed. The contributions of the bbo are already 
extracted, so that the data show the predicted GOE-behavior.}
\end{figure}

For the $\Sigma^2$-statistics which measures long-range correlations the 
effects of the bbo's are strikingly visible, see Fig.~5. Their presence 
influences the rigidity of the spectrum for large values of length $L$. A 
proper handling of these orbits as described for the cumulative NND changes 
the form of the curve towards the expected GOE-like behavior. In the upper 
part of Fig.~5 the experimental data ($\Box$) and the numerical data 
($\triangle$) are displayed for the $\Sigma^2$-statistics before removing the 
contribution of the bbo's. Comparing the $\Sigma^2$-statistics of the 
numerical and the experimental data, no significant difference between both 
data sets can be seen. In contrast to this a small difference between both is 
extracted when the bbo contribution is removed (lower part of Fig.~5). However,
in both cases the $\Sigma^2$-statistics follow very closely the GOE prediction 
up to $L\approx 3.5$, where the distribution saturates. This is in a very good 
agreement with the theoretical predicted value according to \cite{berry} of 
\mbox{$L_{max}\approx 3.8$}. This length $L=L_{max}$ also refers to the 
shortest periodic orbit, which is in the present billiard the bbo. Why is 
$\Sigma^2(L)$ different for $L>L_{max}$ for the two data sets? This might be 
due to the fact that the fabricated microwave cavity is a real system with 
small but existing mechanical imperfections, e.g. a slight non-parallelity of 
the straight segments of the cavity, which should have its effect on the 
periodic orbits. 

After comparing the statistical measures of the two investigated $\gamma=1$
stadium billiards, we now consider their periodic orbits which are given 
through the Fourier transformed spectrum of the eigenvalue density 
$\rho^{fluc}(k)$, see Eq.~(\ref{ft}). The lengths of the classical periodic 
orbits (po) correspond to the positions and their stability roughly spoken to 
the height of the peaks in the spectrum. In Fig.~6 the mod-squared of the 
Fourier transformed of $\rho^{fluc}(k)$ for the experimental (upper part) and 
the numerical (lower part) data are compared. The numerical length spectrum is 
scaled to the experimental one in such a way that the lowest bbo occurs at the 
same location in the two Fourier spectra. Then the bbo's, because of their 
dominating nature in the spectra, were removed \cite{footnote}. Now a direct 
comparison between experiment and simulation becomes possible. It can be seen 
in Fig.~6 that beside the positions of the peaks (lengths of the po) also 
their heights (stability of the po) are almost identical in both spectra.

\begin{figure} [hbt]
\centerline{\epsfxsize=8.6cm
\epsfbox{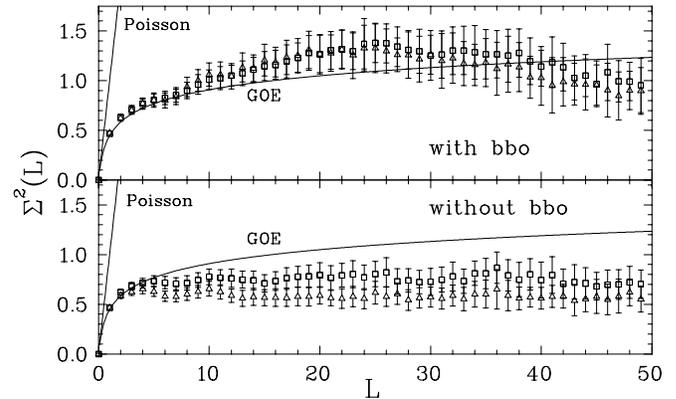}}
\vspace*{2.5ex}
\caption{$\Sigma^2$-statistics of the experimental data set ($\Box$) compared
with the numerical data ($\triangle$). In the upper part the contribution of 
the bbo is not extracted, in the lower part it is. The error bars of the data 
points are a measure of the statistical fluctations within the given ensemble 
of eigenvalues in an interval of length $L$ (see Eq.~(\ref{sigma2formula})). 
Note, the predicted saturation observed in both curves happens at 
$L=L_{max}\approx 4$.}
\end{figure}

\begin{figure} [hbt]
\centerline{\epsfxsize=8.6cm
\epsfbox{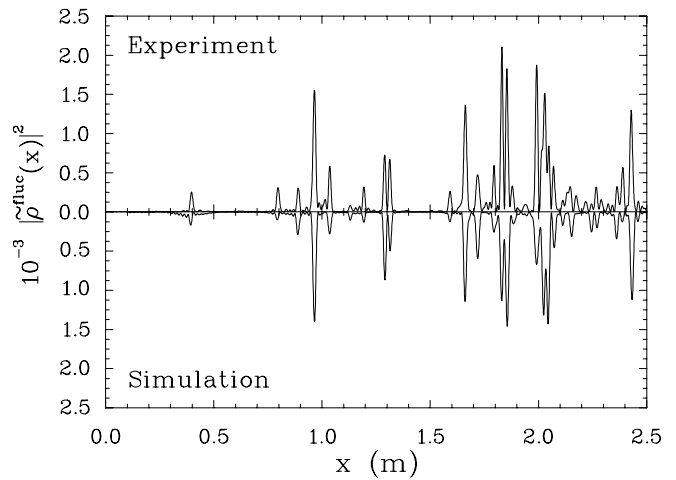}}
\vspace*{2.5ex}
\caption{Fourier transform of the fluctuating part of the level density
$\rho (k)-\rho^{Weyl}(k)-\rho^{bbo}(k)$. The experimental and the numerical 
results are displayed as mirror images. Remnants of the bbo's at $x=0.4$ m, at 
0.8 m and to a lesser extent at 1.2 m are still visible.}
\end{figure}

Let us finally in this subsection return to the bouncing ball orbits. A 
detailed analysis of those in the experimental billiard, which have lengths 
$x=n\cdot R$ \mbox{($n=2,4,6,\ldots$)} shows that the distance $R$ between the 
straight lines of the microwave resonator is not exactly  \mbox{$R=20$ cm} as 
specified in the construction but rather \mbox{$R=(19.92\pm 0.05)$ cm} 
(average over the first six recurrences of the bbo's). That means that the two 
straight segments of the billiard are not exactly parallel. The reasons for 
this deviation ($\Delta R/R=4\times 10^{-3}$) from the design value are on the 
one hand a finite mechanical tolerance during the fabrication ($\approx 
3\times 10^{-3}$) and on the other hand effects coming from thermal 
contractions \cite{erfling} at low temperatures ($\approx 1\times 10^{-3}$). 
This small mechanical imperfection is also the reason for the higher 
saturation level of the $\Sigma^2$-statistics in the experimental data in 
comparison to the numerical one after removing the bbo contribution (lower 
part of Fig.~5). 

\subsection{Influence of the bouncing ball orbits in different Bunimovich 
stadiums}
\label{results_c}
In this section we discuss finally the influence of the bouncing ball orbits 
in the length spectra and in the level statistics of {\it different} Bunimovich 
stadium billiards. Therefore we compare the $\gamma=1$ stadium discussed so 
far with an also measured superconducting $\gamma=1.8$ stadium billiard 
($a=36$ cm and $R=20$ cm), already presented in \cite{graefprl_92}. For this 
comparison we use the first 1060 eigenfrequencies up to a frequency of 18 GHz.

\begin{figure} [hbt]
\centerline{\epsfxsize=8.6cm
\epsfbox{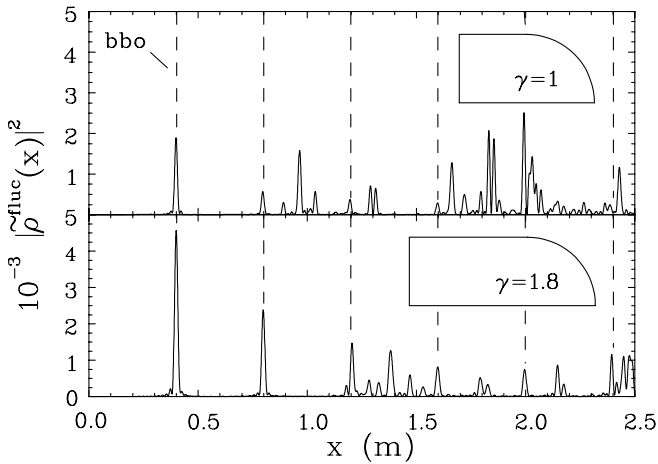}}
\vspace*{2.5ex}
\caption{Experimental length spectra for the $\gamma=1$ (upper part) and the 
$\gamma=1.8$ (lower part) stadium billiard. In both cases the bbo are not 
extracted. Obviously the spectra reveal different amplitudes at the positions 
of the bbo, which are marked by the dashed lines 
($x=n\cdot R,\ n=2,4,6\ldots$).} 
\end{figure}

The interesting features for comparing the two stadiums are the amplitudes of 
the peaks for the bbo and not the peaks of the unstable po's, since they 
naturally have to differ. In Fig.~7 the length spectra of the $\gamma=1$ and 
the $\gamma=1.8$ stadium billiards are shown. It is clearly visible that the 
peaks of the bbo's have different heights or amplitudes in the two 
investigated stadiums. Because the height of the peaks corresponds to the 
stability of the periodic orbits, these differences can be easily explained. 
For the $\gamma=1.8$ stadium the geometrical area on which the bbo's exist is 
almost twice the area of the $\gamma=1$ billiard. This correlation between the 
area of the rectangular part of the billiards with radius $R$ fixed and the 
amplitude of the peaks for the bbo in the FT has been first pointed out in 
\cite {shudo}. This effect is also expressed through the ratio $a/R$ in 
Eq.~(\ref{n_bbo}). 

\begin{figure} [hbt]
\centerline{\epsfxsize=8.6cm
\epsfbox{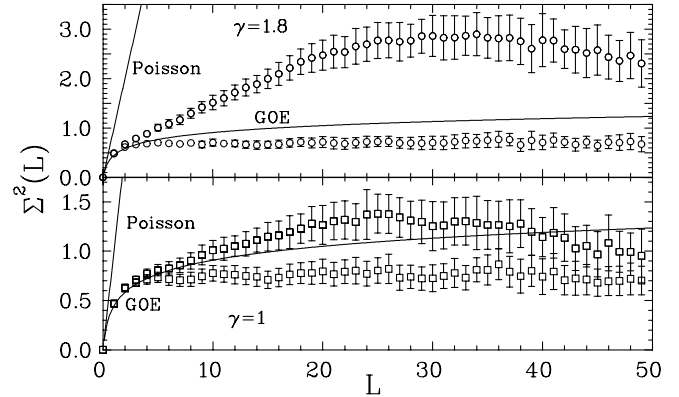}}
\vspace*{2.5ex}
\caption{Comparison of the $\Sigma^2$-statistics of the $\gamma=1.8$ stadium 
($\circ$) before (upper curve)  and after (lower curve)  the extraction of the 
bbo's contribution, as well as for the $\gamma=1$ stadium ($\Box$).}
\end{figure}

The effects of the bbo's are furthermore visible in the statistical measures,
as pointed out in the previous section. For the nearest neighbour spacing 
distribution we have found a Berry-Robnik mixing parameter $q=0.97\pm 0.02$ 
for the experimental $\gamma=1$ billiard before and $q=0.98\pm 0.01$ after 
removing the bbo contribution. But for the $\gamma=1.8$ stadium, presented in 
\cite{graefprl_92}, the situation differs. There we have a mixing parameter 
$q=0.87\pm 0.03$ for the billiard with bbo's and $q=0.97\pm 0.02$ without them.
Using the $\Sigma^2$-statistics to investigate the long range correlations, the 
effect of the bbo's for stadiums of different $\gamma$ values becomes even 
more obvious. The $\Sigma^2$-statistics for the $\gamma=1.8$ stadium, 
including bbo's, increases more strongly, see upper part of Fig.~8 ($\circ$), 
than for the $\gamma=1$ billiard in the lower  part of the same Figure ($\Box$).
After removing the bbo contribution one notices saturation sets in at the same 
$L_{max}\approx4$ for both stadiums. In other words, the shortest po is the 
same in both stadiums as it should be. The results which have been found are 
thus in excellent agreement as predicted in \cite{shudo,bene} which state that 
the influence of the bbo's is in the $\gamma=1$ stadium billiard is very small,
so that it can be characterized as being the most chaotic stadium.

\section{Conclusion}
\label{conclusion}
In this paper we have shown two different methods, an experimental and a 
numerical one, to obtain eigenvalue sequences of a $\gamma=1$ Bunimovich 
stadium billiard. A direct comparison of these two data sets clearly reveals 
that both sets are complete and that no eigenmode is missing. Informations, 
e.g. accuracy of the measured resp. simulated eigenvalues can not be obtained 
from this test. Therefore we analyzed the statistical properties of the 
billiards. Using the cumulative nearest neighbour spacing distribution $I(s)$ 
both data sets show the same predicted GOE behavior after removing the 
contribution of the bouncing ball orbits. The same result is obtained from the 
$\Sigma^2$-statistics, by which we investigated the long range correlations of 
the eigenvalues. Only the saturation level of the experimental data at large 
intervals is somewhat above of the level of the numerical data, because the 
bbo's can not be removed exactly. The reason can be found in the length 
spectra of the data which uncovers that the lengths of the bbo's in the 
experiment slightly deviate from their expected values due to the design of 
the cavity. The cavity contracts at low temperatures and its mechanical 
fabrication has only been possible within certain tolerances. On the other 
hand we have shown that such small imperfections of the real system do not 
have any influence on the results of the statistical analysis. Furthermore, we 
have investigated the influence of the bbo in two different stadium billiards.
They have a clearly different effect on the respective statistical measures,
confirming in particular that the $\gamma=1$ stadium is even more chaotic 
than the $\gamma=1.8$ one.

The comparison of the experimental and the numerical eigenvalues was performed
to show how accurate the results of the statistical analysis depend on the 
given method of providing eigenvalues. To obtain informations about simple 
two-dimensional billiards, like the presented $\gamma=1$ stadium billiard, 
numerical calculations sometimes have an advantage, e.g. if one is interested 
only in eigenvalues. To simulate experiments involving eigenfunctions like in 
\cite{alt_coupling,altprl_95} is a different matter. Furthermore, for problems 
where one is interested in billiards with scatters inside, billiards with 
fractal boundaries or three-dimensional billiards, etc. the experiment clearly 
offers a very convenient way to obtain large sets of eigenvalues quickly.

\section{Acknowledgement}
For the precise fabrication of the niobium resonators we thank H. Lengeler and 
the mechanical workshop at CERN/Geneva. This work has been supported partly 
still by the Sonderforschungsbereich 185 ``Nichtlineare Dynamik'' of the 
Deutsche Forschungsgemeinschaft (DFG) and by the DFG contract RI 242/16-1.

\end{multicols}


\begin{references}
\bibitem {footnote} This, however, was not completely possible -- as can be 
                    seen from the remnants of the lowest bbo in Fig.~6 at 
                    $x=0.4$ m -- due to the numerical uncertainties when 
                    Eq.~(\ref{n_bbo}) is applied to the data.
\end{references}
\end{document}